\documentclass[%
 reprint,
superscriptaddress,
 amsmath,amssymb,
 aps,
]{revtex4-1}

\usepackage{braket}
\usepackage{graphicx}
\usepackage{dcolumn}
\usepackage{bm}
\usepackage{comment}
\usepackage{amsthm}
\theoremstyle{definition}
\newtheorem{theorem}{Theorem}[]

\newtheorem*{theorem*}{Theorem 1}

\begin{document}

\preprint{APS/123-QED}

\title{Breakdown of Markovianity by interactions \\ in stroboscopic Floquet-Lindblad  dynamics under high-frequency drive
}

\author{Kaoru Mizuta}
 \email{mizuta.kaoru.65u@st.kyoto-u.ac.jp}
\affiliation{%
 Department of Physics, Kyoto University, Kyoto 606-8502, Japan
}%
\author{Kazuaki Takasan}%
\affiliation{%
 Department of Physics, University of California, Berkeley, California 94720, USA
}%
\author{Norio Kawakami}
\affiliation{%
 Department of Physics, Kyoto University, Kyoto 606-8502, Japan
}%

\date{\today}

\begin{abstract}
Floquet-Magnus (FM) expansion theory is a powerful tool in periodically driven (Floquet) systems under high-frequency drives. In closed systems, it dictates that their stroboscopic dynamics under a time-periodic Hamiltonian is well captured by the FM expansion, which gives a static effective  Hamiltonian. On the other hand, in dissipative systems driven by a time-periodic Liouvillian, it remains an important and nontrivial problem whether the FM expansion gives a static Liouvillian describing continuous-time Markovian dynamics, which we refer to as the Liouvillianity of the FM expansion. We answer this question for generic systems with local interactions. We find that, while noninteracting systems can either break or preserve Liouvillianity of the FM expansion, generic few-body and many-body interacting systems break it under any finite drive, which is essentially caused by propagation of interactions via higher order terms of the FM expansion. Liouvillianity breaking implies that Markovian dissipative Floquet systems in the high-frequency regimes do not have static (Markovian) counterparts, giving a signature of emergent non-Markovianity. Our theory provides a useful insight for questing unique phenomena in dissipative Floquet systems.
\end{abstract}

\pacs{Valid PACS appear here}
\maketitle


\textit{Introduction.}---Periodically driven (Floquet) systems have attracted much interest as one of the most important class of nonequilibrium systems, which host unique phases such as Floquet topological phases \cite{Kitagawa2010,Rudner2013,Po2016} and time crystals \cite{Sacha2015,Else2016,Keyserlingk2016,Khemani2016,Choi2017,Zhang2017}, and enable controls of the phases (Floquet engineering) \cite{Oka2009, Bukov2015, Oka2019}. 
In particular, Floquet systems in high-frequency regimes, where their frequency $\omega = 2\pi /T$ ($T$: period) is much larger than their energy scale $J$, have been vigorously studied. In closed Floquet systems under a time-periodic Hamiltonian $H(t)$, we can analyze their behavior in these regimes in a unified way by the Floquet-Magnus (FM) expansion, which is a perturbation theory in  $J/\omega$ \cite{Eckardt2015,Bukov2015,Mikami2016}. Importantly, the FM effective Hamiltonian, which approximately describes the stroboscopic dynamics, is a static Hamiltonian, and hence such systems are understood by conventional techniques in static closed systems, leading to Floquet engineering \cite{Oka2009}, Floquet prethermalization \cite{Abanin2017B,Abanin2017Mat,Kuwahara2016,Mori2016} using eigenstate thermalization hypothesis (ETH) \cite{Deutsch1991,Srednicki1994,Rigol2008}, and so on. 
While the hermiticity of the FM effective Hamiltonian provides a powerful tool, it in turn indicates that closed Floquet systems in high-frequency regimes always have counterparts in closed static systems. 

Recently, both theoretical and experimental interest has been spreading out over dissipative Floquet systems \cite{Prosen2011,Szczygielski2014,Shirai2015,Shirai2016,Hartmann2017,Gong2018,Scopa2018,Scopa2019,Gambetta2019,Szczygielski2019,Schnell2020}, with rapidly developing atomic, molecular, and optical platforms \cite{Zhang2017,Tomita2017,Xiao2017,Li2019}. Under Markovianity, dissipative Floquet systems obey the Lindblad equation $\partial_t \rho = \mathcal{L}(t) \rho$ with $\mathcal{L}(t) = \mathcal{L}(t+T)$, where a time-periodic Liouvillian $\mathcal{L}(t)$ is given by
\begin{equation}\label{FloquetLiouvillian}
\mathcal{L}(t) \rho = -i [H(t),\rho] + \sum_i L_i(t) \rho L_i(t)^\dagger - \frac{1}{2} \{ L_i(t)^\dagger L_i(t), \rho \}.
\end{equation}
The linearity of $\mathcal{L}(t)$ implies a possible extension of the FM expansion to dissipative systems, and a static linear operator $\mathcal{L}_f^{n}$ [defined by Eq.~(\ref{FMeffective0}) below] called the FM effective Lindbladian is obtained. 
However, in contrast to closed systems, it is an important and nontrivial problem whether the FM effective Lindbladian $\mathcal{L}_f^n$ is a static Liouvillian given by the time-independent version of Eq. (\ref{FloquetLiouvillian}), which we call Liouvillianity. Liouvillianity of a static super-operator ensures a completely-positive and trace-preserving (CPTP) time-homogeneous evolution under Markovianity, providing some important properties of Marikovian open systems \cite{Lindblad1976,Breuer2002,Rivas2012}. For instance, it ensures the existence of a nonequilibrium steady state (NESS) and the validity of the trajectory method \cite{Dalibard1992, Dum1992, Rivas2012, Daley2014}. Liouvillianity of $\mathcal{L}_f^n$ also determines whether Markovian dissipative Floquet systems in high-frequency regimes are understood as Markovian dissipative static systems. 

Recent studies on high-frequency regimes of dissipative Floquet systems revealed their approximate  dynamics \cite{Dai2016} and NESS \cite{Ikeda2020}, but Liouvillianity itself was not focused on. While Ref. \cite{Haddadfarshi2015} first evaluated Liouvillianity of the FM expansions, these previous studies mainly focused on noninteracting systems. Thus, the knowledge of Liouvillianity of the FM expansions is still lacking in generic systems, especially in interacting systems.

Here, we address the fundamental question \textit{whether the FM effective Lindbladian is a Liouvillian} in generic systems with particular emphasis on  few-body or many-body systems with local interactions. We find out that, while noninteracting systems show model-dependent behaviors, breaking or preservation of Liouvillianity, generic interacting systems experience Liouvillianity breaking of the FM expansions under any finite drive. Importantly, Liouvillianity breaking in interacting systems is essentially attributed to a spreading structure of local interactions in FM expansions. Liouvillianity breaking implies that such Floquet systems cannot be captured as static Markovian systems: namely dissipative Floquet systems under Markovianity can show emergent non-Markovianity in stroboscopic dynamics. Our theory can be used for studying unique phenomena in dissipative interacting Floquet systems.

\textit{Floquet-Magnus effective Lindbladian.}---First, we briefly introduce the FM expansion for dissipative Floquet systems \cite{Dai2016}. Under a time-periodic Liouvillian $\mathcal{L}(t)$ [Eq. (\ref{FloquetLiouvillian})], we define the effective Lindbladian $\mathcal{L}_\text{eff}$ and the Floquet operator $\mathcal{U}_\text{eff}$ by
\begin{equation}\label{FloquetOp}
\mathcal{L}_\text{eff} = \frac{1}{T} \log \mathcal{U}_\text{eff}, \quad \mathcal{U}_\text{eff} = \mathcal{T} \exp \left( \int_0^T \mathcal{L}(t)dt\right).
\end{equation}
The FM effective Lindbladian $\mathcal{L}_f^n$ is obtained by the perturbative expansion for $\mathcal{L}_\text{eff}$ up to the $n$-th order of $||\mathcal{L}(t)||_\text{op}/\omega$ ($||\quad ||_\text{op}$: operator norm), and this results in 
\begin{eqnarray}
\mathcal{L}_f^{n} &=& \sum_{i=0}^n \mathcal{L}_f^{(i)}, \quad \mathcal{L}_f^{(0)} = \frac{1}{T} \int_0^T \mathcal{L}(t) dt, \label{FMeffective0} \\
\mathcal{L}_f^{(1)} &=&  \frac{1}{2T} \int_0^T dt_1 \int_0^{t_1} dt_2 [\mathcal{L}(t_1),\mathcal{L}(t_2)]. \label{FMeffective12}
\end{eqnarray}
This perturbative expansion has the same convergence radius as the one for closed systems, and hence $\mathcal{L}_f^n$ well captures the stroboscopic dynamics at $t=mT$ $(m \in \mathbb{N})$ within $||\mathcal{L}(t)||_\text{op}/\omega < O(1)$ \cite{Bukov2015,Sconvergence}. Since Liouvillians are closed only with respect to the summation, Liouvillianity of $\mathcal{L}_f^{n}$ and each $i$-th order term $\mathcal{L}_f^{(i)}$ are nontrivial, and to be clarified.

\textit{Condition for Liouvillianity.}---A super-operator $\mathcal{L}$ on states $\rho$ is called a Liouvillian if its time evolution operator $\exp (\mathcal{L} t)$ becomes a CPTP map for $\,^\forall t \geq 0$. It is equivalent to that $\mathcal{L}$ is given by the Lindblad form, which is the time-independent version of Eq. (\ref{FloquetLiouvillian}).  We refer to whether $\mathcal{L}$ is a Liouvillian as Liouvillianity of $\mathcal{L}$. Here, we introduce some mathematical tools and describe how to judge Liouvillianity of the FM expansions.

We denote a set of $d \times d$ matrices by $\mathbb{M}_d$ and assume a state $\rho \in \mathbb{M}_d$. We define the Frobenius basis $\{ F_j \}_{j=1}^{d^2}$ as a complete orthonormal set (CONS) for $\mathbb{M}_d$. Using the Frobenius inner product $\left< A, B \right>_\text{F}= \mathrm{Tr}[A^\dagger B]$, it satisfies the orthonormality relations $\left< F_j, F_k \right>_\text{F} = \delta_{jk}$, $\mathrm{Tr} [F_j] = 0$ if $j \neq d^2$, and $F_{d^2} = I/\sqrt{d}$. Next, we introduce the doubled Hilbert space representation~\cite{Jamiolkowski1973,Choi1975}, in which we regard a state $\rho=(\rho)_{ij} \in \mathbb{M}_d$ as a $d^2$-dimensional vector $\vec{\rho}=(\rho)_{(ij)}$. Then, any linear operator on $\rho$ is represented by a matrix in $\mathbb{M}_{d^2}$, and a Liouvillian $\mathcal{L}$ is given by
\begin{eqnarray}
\mathcal{L} &=& -i (H \otimes I - I \otimes H^\mathrm{T}) \nonumber \\
&&+ \sum_{j,k=1}^{d^2-1} a_{jk} \left[ F_j \otimes F_k^\ast - \frac{1}{2} ( F_k^\dagger F_j \otimes I + I \otimes F_j^\mathrm{T} F_k^\ast )\right]. \nonumber \\
\label{LiouvillianDoubled} &&
\end{eqnarray}
with a hermitian matrix $H \in \mathbb{M}_d$ (Hamiltonian) and a hermitian positive-semidefinite matrix $[a_{jk}]_{j,k=1}^{d^2-1} \in \mathbb{M}_{d^2-1}$ (dissipator). In this representation, the system size becomes double, and an action $A \rho B$ ($A,B \in \mathbb{M}_d$) is written as $A \otimes B^{\mathrm{T}}$. We call the system, which $A$ ($B^\text{T}$) acts on, a real (fictitious) system.

Using the hermiticity-preserving property $(\mathcal{L}(t)[\rho])^\dagger=\mathcal{L}(t)[\rho^\dagger]$ ($\,^\forall \rho \in \mathbb{M}_d$) and the trace-preserving property $\mathrm{Tr}(\mathcal{L}(t)[\rho])=0$ ($\,^\forall \rho \in \mathbb{M}_d$), the $n$-th order FM expansion $\mathcal{L}_f^n$ is always written in the same form as Eq. (\ref{LiouvillianDoubled}) for any $n$
with hermitian matrices $H = H^n \in \mathbb{M}_d$ and  $[a_{jk}]_{j,k=1}^{d^2-1}  = [a_{jk}^n]_{j,k=1}^{d^2-1} \in \mathbb{M}_{d^2-1}$ (See Ref. \cite{Haddadfarshi2015} or  Sec. S1 of Supplemental Material \cite{Sfm} for information about the FM expansion and the derivation of our results on its properties). Note that $[a_{jk}^n]$ is not always positive-semidefinite, and hence $\mathcal{L}_f^n$ is not always a Liouvillian. Using the orthonormality of the Frobenius basis, the condition for Liouvillianity is summarized as follows:
\begin{eqnarray}
&& \text{$\mathcal{L}_f^n$ is a Liouvillian} \nonumber \\ 
&& \Leftrightarrow \text{$[a_{jk}^n]_{j,k=1}^{d^2-1} \in \mathbb{M}_{d^2-1}$ is positive-semidefinite,} \label{ConditionLiouvillianity}
\end{eqnarray}
with
\begin{equation} \label{a_jk}
a_{jk}^n = \text{Tr}[(F_j^\dagger \otimes F_k^\text{T}) \mathcal{L}_f^n].
\end{equation}
When $[a_{jk}^n]$ has negative eigenvalues, $\mathcal{L}_f^n$ breaks Liouvillianity. Then, we define the degree of Liouvillianity breaking by the smallest negative eigenvalue of $[a_{jk}^n]$.  Considering the expected accuracy $\mathcal{L}_{\text{eff}} = \mathcal{L}_f^n + O((||\mathcal{L}(t)||/\omega)^{n+1})$, this degree measures how the effect of Liouvillianity breaking of $[a_{jk}^n]$ appears in the real-time dynamics.

Each $i$-th order term $\mathcal{L}_f^{(i)}$ is also written in the same form as Eq. (\ref{LiouvillianDoubled}), and thus we can judge its Liouvillianity in the same way. However, we obtain the following result (see Supplemental Material \cite{Sfm}):
\begin{flushleft}
(a) $\mathcal{L}_f^{(0)}$ ($= \mathcal{L}_f^0$) is always a Liouvillian. \\
(b) $\mathcal{L}_f^{(i)}$ ($i \geq 1$) is a Liouvillian if and only if
\begin{equation}\label{LiouvillianityTrivial}
[a_{jk}^{(i)}]=\left[ \mathrm{Tr} [ (F_j^\dagger \otimes F_k^\mathrm{T}) \, \mathcal{L}_f^{(i)}] \right] = O_{d^2-1},
\end{equation}
where $O_d$ ($\in \mathbb{M}_d$) represents a zero matrix with the size $d$.
\end{flushleft}
Thus, except for special cases where $\mathcal{L}_f^{(i)}$ gives no dissipation, any higher order term $\mathcal{L}_f^{(i)}$ $(i \geq 1)$ is not a Liouvillian in general. This brings an essential difference from closed systems. In closed systems, each order term $H_f^{(i)}$ is always a Hamiltonian, and hence the FM effective Hamiltonian $H_f^n = \sum_{i=0}^n H_f^{(i)}$ is also a Hamiltonian. On the other hand, in dissipative cases, generic higher order terms $\mathcal{L}_f^{(i)}$ $(i \geq 1)$ do harm to the Liouvillianity of $\mathcal{L}_f^n = \sum_{i=0}^n \mathcal{L}_f^{(i)}$. 

Considering such a property of each order term $\mathcal{L}_f^{\left(i\right)}$, we discuss whether Liouvillianity of the FM effective Lindbladian $\mathcal{L}_f^n$ is preserved or broken in noninteracting and interacting systems below. Before the discussion, we note that Liouvillianity breaking of $\mathcal{L}_f^n$ is not necessarily unphysical, although it seemingly breaks complete-positivity. This is because the time evolution operator generated by the effective Lindbladian is meaningful only at discrete time while Liouvillianity is just a condition for CPTP dynamics at any continuous time $t \geq 0$. In fact, Refs. \cite{Hartmann2017,Schnell2020} also numerically observed the Liouvillianity breaking of the effective Lindbladian [Eq. (\ref{FloquetOp})] in noninteracting systems. Reference \cite{Dai2016} found that the FM expansion well describes the stroboscopic dynamics, where it breaks Liouvillianity (Liouvillianity itself is not discussed in Ref. \cite{Dai2016}, but we confirm its breaking).

\textit{Liouvillianity in noninteracting systems.}--- We first discuss Liouvillianity of the FM effective Lindbladian for noninteracting systems, and show that noninteracting systems host two model-dependent phenomena. We focus on a single spin with $S=1/2$, and the Frobenius basis is given by the Pauli matrices $\sigma^i$:
$F_i = \sigma^i / \sqrt{2}~(i=1,2,3), F_4 = \sigma^0 / \sqrt{2}$.
We consider a time-periodic drive (the period $T=2\tau > 0$):
\begin{equation}\label{ModelNoninteracting}
\mathcal{L}(t) \rho = \begin{cases}
-i h [\sigma^3,\rho] + \gamma_2 ( \tilde{\sigma}^{13}\rho \tilde{\sigma}^{13} -\rho ) & (0\leq t < \tau) \\
\gamma_1 (\sigma^1 \rho \sigma^1 - \rho) & ( \tau \leq t < 2\tau ),
\end{cases}
\end{equation}
with $\tilde{\sigma}^{13} =  (\sigma^1 + \sigma^3 )  / \sqrt{2}$. Here, $h \in \mathbb{R}$ is the strength of the magnetic field in  the $z$-direction, and the parameters $\gamma_1 (> 0)$ and $\gamma_2 (\geq 0)$ represent dephasing in certain directions. We assume the high-frequency regime where the frequency $\omega=\pi/\tau$ is much larger than $h$, $\gamma_1$, and $\gamma_2$, or equivalently we assume $h\tau, \gamma_1 \tau, \gamma_2 \tau \ll 1$. We discuss two different models $\mathcal{L}_A(t)=\mathcal{L}(t)|_{h \neq 0, \gamma_2=0}$ and $\mathcal{L}_B(t)=\mathcal{L}(t)|_{h=0, \gamma_2 >0}$ and evaluate the FM expansion up to the second order. For the first model $\mathcal{L}_A(t)$, we obtain
\begin{equation}\label{ModelA_second}
[a_{jk}^{2}] = \gamma_1 \left(\begin{array}{ccc}
1- \alpha_A^{(2)} & \alpha_A^{(1)} & 0 \\
 \alpha_A^{(1)} & \alpha_A^{(2)} & 0 \\
0 & 0 & 0
\end{array}\right), 
\end{equation}
with $\alpha_A^{(1)}= - h \tau$ and  $\alpha_A^{(2)} = 2(h \tau)^2/3$. We also obtain the result for the second model $\mathcal{L}_B(t)$ as
\begin{equation}\label{ModelB_second}
[a_{jk}^{2}] = \left(\begin{array}{ccc}
\gamma_2/2 + \gamma_1 + \alpha_B^{(2)} \gamma_2 & 0 & \gamma_2/2 + \alpha_B^{(2)} \gamma_1 \\
0 & 0 & 0 \\
\gamma_2/2 + \alpha_B^{(2)} \gamma_1 & 0 & \gamma_2/2 - \alpha_B^{(2)} \gamma_2
\end{array}\right)
\end{equation}
with $\alpha_B^{(2)}=(\gamma_1 \tau)(\gamma_2 \tau)/6$. $\alpha_A^{(1)}$ and $\alpha_{A,B}^{(2)}$ represent the first and the second order terms (the first order term vanishes in the second model). Setting them to zero properly, the zeroth and first order results $[a_{jk}^0]$ and $[a_{jk}^1]$ are reproduced.

By evaluating the positive-semidefiniteness of $[a_{jk}^n]$ $(n \leq 2)$, we observe two different phenomena. For the first model $\mathcal{L}_A(t)$, the first-order FM expansion $\mathcal{L}_f^1$ is not a Liouvillian under infinitesimal $\alpha_A^{\left(1\right)} $since the matrix $[a_{jk}^n]$ always possesses a negative eigenvalue $(1-\sqrt{1+4(h\tau)^2})\gamma_1/2\simeq-\gamma_1\left(h\tau\right)^2=O((h\tau)^2)$. Thus, while $\mathcal{L}_f^1$ always breaks Liouvillianity, the degree of Liovillianity breaking, measured by the negative eigenvalue of $[a_{jk}^n]$, is small when we consider the accuracy of the FM expansions. Similar phenomenon has been observed in Ref. \cite{Haddadfarshi2015}, in which they construct a new FM effective Lindbladian recovering Liouvillianity in such situations. At the second order for the first model, the matrix $[a_{jk}^2]$ always possesses a negative eigenvalue
$(1-\sqrt{1+4(h\tau)^2/3+16(h\tau)^4/9})\gamma_1/2\simeq-\gamma_1 (h\tau)^2/3=O((h\tau/\omega)^2)$,
and hence $\mathcal{L}_f^2$ always breaks Liouvillianity. Different from the first order, the degree of Liouvillianity breaking is large enough, and it affects the real-time dynamics to non-negligible extent in the time scale where $\mathcal{L}_f^2$ becomes valid. Including higher orders in which Liouvillianity breaking with the degree $O((h\tau/\omega)^2)$ remains, the first model $\mathcal{L}_A(t)$ always breaks Liouvilianity of the FM expansions.

On the other hand, for the second model $\mathcal{L}_B(t)$, the smallest eigenvalue of the matrix $[a_{jk}^2]$ is the smaller value of $0$ and
\begin{equation}
\frac{\gamma_1+\gamma_2}{2}-\frac{1}{6}\sqrt{\gamma_1^2\gamma_2^2\tau^2 ( \gamma_1^2 \tau^2 + \gamma_2^2 \tau^2 +12)+9\left(\gamma_1^2+\gamma_2^2\right)}.
\end{equation}
This matrix is positive-semidefinite only within the range $0<\tau\leq \tau_{\mathrm{max}}$, where $\tau_{\mathrm{max}}$ is given by
\begin{equation}
\tau_{\mathrm{max}}=\left[\frac{6\left(\sqrt{1+\left(\gamma_1^2+\gamma_2^2\right)/\left(2\gamma_1\gamma_2\right)}-1\right)}{\gamma_1^2+\gamma_2^2}\right]^{1/2}>0.
\end{equation}
In other words, the second order FM expansion $\mathcal{L}_f^2$ preserves Liouvillianity within this finite parameter range $0<\tau\le\tau_{\mathrm{max}}$.

\begin{figure*}
\hspace{-1cm}
\begin{center}
    \includegraphics[height=4.5cm, width=18cm]{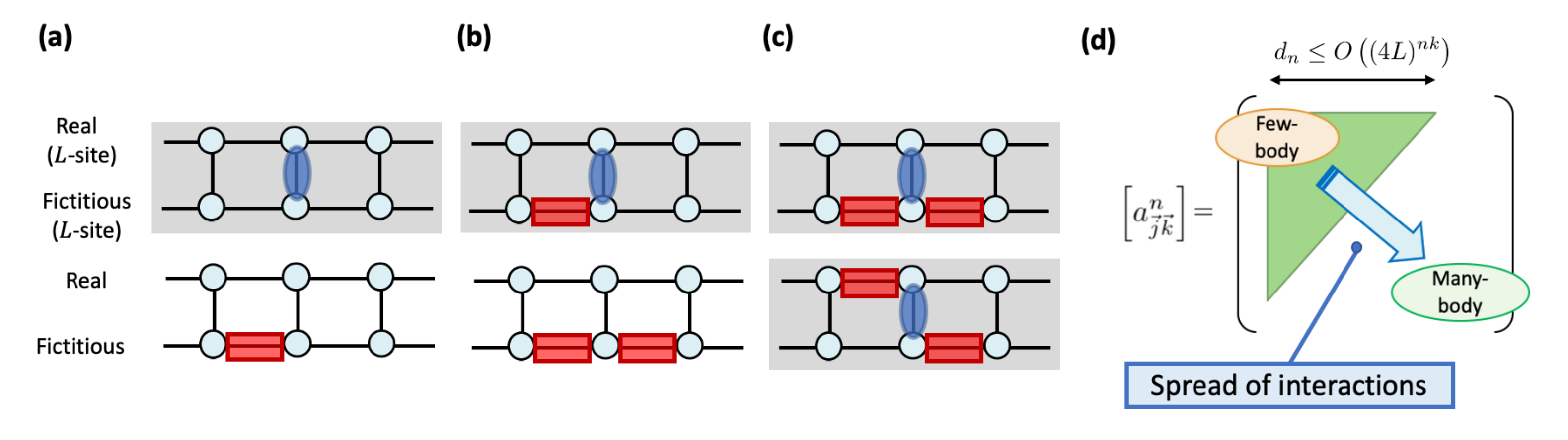}
    \caption{(a,b,c) Some of local terms appearing in the FM expansions with the locality $k=2$. The zeroth-, first-, and second-order terms are described by (a), (b), and (c) respectively. Local terms designated by  red rectangles (closed within either a real chain or a fictitious one) and  blue ellipses (connecting both chains) are caused by the Hamiltonian $H(t)$ and dissipation $L_i(t)$ respectively. Local terms with gray backgrounds, involving both real and fictitious systems, appear in $[a_{\vec{j}\vec{k}}^n]$. (d) Generic form of $[a_{\vec{j}\vec{k}}^n]$. It reflects the spread of local interactions in (a), (b), and (c).} 
    \label{Figure}
 \end{center}
 \end{figure*}

\textit{Extension to interacting models.}---We now discuss interacting few-body and many-body systems. We consider an $L$-site spin chain with $S=1/2$, and the Frobenius basis is a set of the $4^L$ matrices in $\mathbb{M}_{2^L}$, $F_{\vec{j}} = \frac{1}{\sqrt{2^L}} \prod_{l=1}^{L} \sigma^{j_l}_l, \vec{j}=(j_1,\hdots,j_l,\hdots,j_L)$
with $j_l=0,1,2,3$. By denoting $\vec{0}=(0,0,\hdots,0)$, Liouvillianity of the FM effective Lindbladian $\mathcal{L}_f^n$ is confirmed when the matrix $[a_{\vec{j}\vec{k}}^n]_{\vec{j},\vec{k} \neq \vec{0}} \in \mathbb{M}_{4^L-1}$, given by $a_{\vec{j}\vec{k}}^n = \mathrm{Tr} [(F_{\vec{j}}^\dagger \otimes F_{\vec{k}}^\mathrm{T} ) \, \mathcal{L}_f^n ]$, is positive-semidefinite.  The difficulty compared to noninteracting systems is that the size of $[a_{\vec{j}\vec{k}}^n]$ is exponentially large in $L$, which will be overcome by locality of interactions below.

We begin with a model driven by Ising interactions and dephasing with the periodic boundary conditions:
\begin{equation}\label{ModelC}
\mathcal{L}_C(t) \rho = \begin{cases}
-i J_z \sum_l [\sigma_l^3 \sigma_{l+1}^3, \rho] \equiv \mathcal{L}_{C1} \rho& ( 0 \leq t < \tau )\\
\gamma \sum_l (\sigma_l^1 \rho \sigma_l^1 - \rho) \equiv \mathcal{L}_{C2} \rho & ( \tau \leq t < 2\tau ).
\end{cases}
\end{equation}
The zeroth order $\mathcal{L}_f^0$ is the time-average $\mathcal{L}_f^0=(\mathcal{L}_{C1}+\mathcal{L}_{C2})/2$, and the first order is
\begin{eqnarray}
\mathcal{L}_f^1 = \mathcal{L}_f^0 &+& \frac{J_z \gamma \tau}{2} \sum_l
\left\{ \sigma_l^1 \otimes  \sigma_l^2 (\sigma_{l-1}^3 + \sigma_{l+1}^3) \right. \nonumber \\
&\qquad& \qquad \qquad \,  \left. -  \sigma_l^2 (\sigma_{l-1}^3 + \sigma_{l+1}^3)  \otimes \sigma_l^1 \right\}. \label{ModelC_first}
\end{eqnarray}
Up to the second order, though the matrix $[a_{\vec{j}\vec{k}}^{2}]$ possesses $O(L)$ nonzero components, we can rewrite it in a block-diagonalized form by arranging the order of the basis:
\begin{equation}\label{ModelC_second}
[a_{\vec{j}\vec{k}}^{2}] =  \left[ \bigoplus_{l=1}^L  \tilde{\gamma} \left( \begin{array}{cccc}
1 - 2\alpha^{(2)} & \alpha^{(1) }& \alpha^{(1) }& -\alpha^{(2)}  \\
\alpha^{(1)} & \alpha^{(2)} & \alpha^{(2)} & 0 \\
\alpha^{(1)} &  \alpha^{(2)} & \alpha^{(2)} & 0 \\
- \alpha^2  & 0 & 0 & 0
\end{array} \right)_l
\right] \oplus O_{D},
\end{equation}
with $\tilde{\gamma}=2^{L-1}\gamma$, $\alpha^{(1)} = - J_z \tau$, $\alpha^{(2)} = 2(J_z \tau)^2/3$, and  $D=4^L -4L -1$. The basis of the $4 \times 4$ matrices $(\cdots)_l$ is composed of $\vec{j}=(\hdots 0,j_l=1,0 \hdots), (\hdots 0,2,j_l=3,0 \hdots),(\hdots 0,j_l=3,2,0 \hdots), (\hdots 0,3,j_l=1,3,0 \hdots)$. 
Up to the first order with $\alpha^{(2)}=0$, the matrix $[a_{\vec{j}\vec{k}}^1]$ has a negative eigenvalue $\gamma \cdot 2^{L-2} \{ 1- \sqrt{1+8(J_z \tau)^2} \}$. Thus, the FM effective Lindbladian $\mathcal{L}_f^1$ always breaks Liouvillianity, though its degree characterized by $O((J_z \tau)^2)$ is small. The matrix $[a_{\vec{j}\vec{k}^2}]$, which is for the second-order, also possesses negative eigenvalues. By numerical calculations, the smallest one $\lambda_{\text{min}}$ is  well fitted by
\begin{equation}
\lambda_{\text{min}}/[\gamma \cdot 2^{L-1} (J_z \tau)^2] \simeq \sum_{m=0}^3 C_m (J_z \tau)^m,
\end{equation}
\begin{equation*}
C_0=-0.667, C_1=0.0197, C_2=-3.08, C_3=2.84,
\end{equation*}
with the root mean square error $3.25\times 10^{-4}$ in $0< J_z \tau < 0.5$. Thus, the second order one $\mathcal{L}_f^{2}$ always breaks Liouvillianity with the degree $O((J_z\tau)^2)$, which is non-negligible. Including higher-order FM expansions $\mathcal{L}_f^n$, where $n$ is much smaller than the system size $L$, this interacting model $\mathcal{L}_C(t)$ shows Liouvillianity breaking of the FM effective Lindbladian regardless of parameters.

We also observe the same behavior in another interacting model (see Sec. S2 in the Supplemental Material \cite{Sfm} for discussion on the other interacting model). In both interacting cases, immediate Liouvillianity breaking for any $\tau > 0$ can be attributed to a spreading structure of $[a_{\vec{j}\vec{k}}^n]$, appearing also in the noninteracting model $\mathcal{L}_B(t)$. However, different from noninteracting cases, higher order terms out of the block where $[a_{\vec{j}\vec{k}}^0]$ lies come from interactions involving a larger number of sites [for example, three-body terms in Eq. (\ref{ModelC_first})], which is proven to be essential in interacting systems.

\textit{Liouvillianity breaking in generic interacting systems.}---We finally show that immediate Liouvillianity breaking of the FM effective Lindbladian takes place in generic few-body and many-body systems with local interactions. Importantly, the above spreading structure of $[a_{\vec{j}\vec{k}}^n]$ universally appears due to the propagation of local interactions in FM expansions, leading to Liouvillianity breaking. Although we discuss an $L$-site spin chain with $S=1/2$ here, our results are easily generalized to any-dimensional finite systems with finite degrees of freedom.


Here, we assume the $k$-locality for interactions, indicating that the Hamiltonian $H(t)$ and the Lindblad operator $L_i(t)$ in Eq. (\ref{FloquetLiouvillian}) include at-most $k$-body and $(k/2)$-body interactions respectively. We also assume that $H(t)$ includes at-least two-body interactions. The complex energy per site $\sim ||\mathcal{L}(t)||_\text{op}/L$ is assumed to be bounded by $J$, which is physically reasonable. For our interacting model [Eq. (\ref{ModelC})], we can take $k=2$ and $J=4J_z+2\gamma$. Under these assumptions, we obtain the following rigorous bound (see Supplemental Material \cite{Sfm}):
\begin{equation}\label{a_jk_bound}
\left|  a_{\vec{j}\vec{k}}^{(i)}  \right| \leq \frac{(2kJT)^i}{i+1} J \cdot i! \cdot 2^L.
\end{equation}
As a result, $a_{\vec{j}\vec{k}}^{(i)}$ decays within lower orders in high-frequency expansion up to the order $n < 1/(2kJT)$. Thus, the problem itself for Liouvillianity does not differ from noninteracting cases where whether $[a_{\vec{j}\vec{k}}^{(0)}]$ can remain positive-semidefinite under perturbations of $[a_{\vec{j}\vec{k}}^{(i)}]$ determines the Liouvillianity. However, the important difference from the noninteracting case is that the locality of interactions restricts the form of $[a_{\vec{j}\vec{k}}^{n}]$. In the doubled Hilbert space representation, a $k$-local Liouvillian $\mathcal{L}(t)$ on an $L$-site system is interpreted as a nonhermitian Hamiltonian with $k$-body interactions on real and fictitious systems which have $L$-sites respectively as shown in Fig.~\ref{Figure}~(a). Under the $k$-locality, the commutator $[\mathcal{L}(t_1),\mathcal{L}(t_2)]$  and thereby $\mathcal{L}_f^{(1)}$ include at-most $(2k-1)$-body interactions, since it is composed of the commutators of local terms in $\mathcal{L}(t)$ which have overlaps on at-least one-site. Considering that generic $i$-th order terms $\mathcal{L}_f^{(i)}$ are composed of $i$-tuple 
multi-commutators of $\mathcal{L}(t)$, the sum $\mathcal{L}_f^{n}$ includes at-most $\{ (n+1)k-n \}$-body interactions. We denote the number of $l$ satisfying $j_l \neq 0$ in $\vec{j}$ by $n_{\vec{j}}$, and then nonzero $a_{\vec{j}\vec{k}}^n$ ensures the existence of a $(n_{\vec{j}}+n_{\vec{k}})$-body term $F_{\vec{j}} \otimes F_{\vec{k}}^\ast$, involving both of real and fictitious systems 
[See Fig.~\ref{Figure}~(a)]. Thus, the locality constraint gives
\begin{equation}\label{a_jk_zero}
a_{\vec{j}\vec{k}}^n = 0 \quad \text{if} \quad (n_{\vec{j}}+n_{\vec{k}}) > \{ (n+1)k-n \}.
\end{equation}
By rearranging the Frobenius basis in ascending order of the locality $n_{\vec{j}}$, the matrix $[a_{\vec{j}\vec{k}}^n]$ is block-diagonalized as follows:
\begin{equation}\label{a_jk_BlockDiagonal}
\left[ a_{\vec{j}\vec{k}}^n \right] = A_{d_n} \oplus O_{4^L-d_n-1}, \quad A_{d_n} \in \mathbb{M}_{d_n}.
\end{equation}
The basis of the nontrivial part $A_{d_n}$ is  composed of $\vec{j}$ with $1 \leq n_{\vec{j}} \leq (n+1)k-n$, and the size $d_n$ satisfies
\begin{equation}\label{d_n}
 d_n \leq \,_L C_{(n+1)k-n} \cdot 4^{(n+1)k-n} \sim \frac{(4L)^{(n+1)k-n}}{\{(n+1)k-n\}!}.
\end{equation}
Furthermore, Eq. (\ref{a_jk_zero}) also indicates that the elements where both $n_{\vec{j}}$ and $n_{\vec{k}}$ exceed $\lceil (n+1)k/2-n/2 \rceil$ vanish ($\lceil x \rceil$: the ceil function). Thus, assuming $d_n < 4^L -1$, the nontrivial part $A_{d_n}$ is always written in the form of
\begin{equation}\label{A_d_n}
A_{d_n} = \left( \begin{array}{cc}
\tilde{A}_{e_n} & \tilde{B} \\
\tilde{B}^\dagger & O_{d_{n}-e_n}
\end{array}\right), \quad
\text{$\tilde{A}_{e_n} \in \mathbb{M}_{e_n}$ : hermitian},
\end{equation}
with $e_n \leq \,_L C_{\lceil (n+1)k/2-n/2 \rceil} \cdot 4^{\lceil (n+1)k/2-n/2 \rceil}$. This triangular form is attributed to the propagation of interactions via higher order terms [Fig. \ref{Figure}(b)], where the Hamiltonian $H(t)$ (the dissipation $L_i(t)$) causes spread closed within real or fictitious systems (over both systems) [Fig. \ref{Figure}(a)]. If the interactions of $H(t)$ and $L_i(t)$ are neighboring (simultaneously acting on at-most the $k$-th and $(k/2)$-th nearest neighbors respectively), the size $d_n$ reduces to $O(4^{(n+1)k-n}L)$.

The increasing dimension $d_n$ with the order $n$ means the spreading structure of $[a_{jk}^n]$ from the zeroth order in common with the models $\mathcal{L}_B(t)$ [Eq. (\ref{ModelNoninteracting})] and $\mathcal{L}_C(t)$ [Eq. (\ref{ModelC})] showing the immediate Liouvillianity breaking. This perturbs zero eigenvalues in $[a_{jk}^0]$ and can shift them to negative. More rigorously, using the Schur complement \cite{Zhang2005}, the triangular hermitian matrix [Eq. (\ref{A_d_n})] always has at-least $(\text{rank} \tilde{B})$ negative eigenvalues, and hence $\mathcal{L}_f^n$ for $n \geq 1$ is always a non-Liouvillian as long as $\tilde{B} \neq O$. We conclude that \textit{Liouvillianity of the FM effective Lindbladian is always broken in generic interacting systems due to the spread of interactions}.

\textit{Discussion.}---Here, we would like to discuss the versatility of our results. Since the essence of Liouvillianity breaking in interacting systems is the spread of interactions through commutators, our discussion is valid for other types of high-frequency expansions such as the Schrieffer-Wolff expansions \cite{Bukov2015}, and the van Vleck expansions given by
\begin{equation}
\mathcal{L}_\text{vV}^n = \sum_{i=0}^n \mathcal{L}_\text{vV}^{(i)}, \, \mathcal{L}_\text{vV}^{(0)} = \mathcal{L}_0, \,
\mathcal{L}_\text{vV}^{(1)} =  \sum_{m=1}^{\infty} \frac{[\mathcal{L}_{-m}, \mathcal{L}_m]}{2m\omega},  \, \hdots \, , 
\end{equation}
where $\mathcal{L}_m$ represents the Fourier component $\mathcal{L}_m = \int_0^T \mathcal{L}(t) e^{-i2\pi mt/T} dt /T$ \cite{Dai2016,Ikeda2020}. These expansions always break Liouvillianity in generic interacting systems as long as the spread of interactions takes place.

We also note that some exceptions do not host Liouvillianity breaking at any order or up to a certain order by avoiding the spread of local interactions. The first exception  is a commutative Liouvillian which satisfies $[\mathcal{L}(t_1),\mathcal{L}(t_2)]=0$ ($\,^\forall t_1, \,^\forall t_2$) \cite{Szczygielski2019}. Then, $\mathcal{L}_f^n = \mathcal{L}_0$ trivially becomes a Liouvillian regardless of interactions at any order $n$. Floquet systems under time-independent dissipation, where $L_i(t)$ in Eq. (\ref{FloquetLiouvillian}) is time-independent, is another exception which preserves Liouvillianity of the van Vleck effective Lindbladians up to the first order \cite{Ikeda2020}, while higher order ones generally break Liouvillianity. This is because the first order term $\mathcal{L}_{\text{vV}}^{(1)}$ gives no dissipation when dissipation is time-independent, or equivalently, there is no spread of interactions in dissipative terms up to the first order. Though we expect that some other specific solvable models can preserve Liouvillianity, they will also experience Liouvillanity breaking under perturbations, which inevitably causes the spread of interactions. 


\textit{Conclusions.}---We have considered dissipative Floquet systems in high-frequency regimes, and have evaluated the Liouvillianity of the FM effective Lindbladian for noninteracting systems and locally interacting systems. While noninteracting systems show model-dependent phenomena,
we have provided interacting models rigorously showing the immediate breakdown of Liouvillianity. We have developed a theoretical framework to judge Liouvillianity breaking in terms of the structure of a hermitian matrix $[a_{jk}^n]$ determined by the FM effective Lindbladian, and have demonstrated that the spread of interactions via higher order terms in generic interacting systems always causes the Liouvillianity breaking of the FM effective Lindbladian.

Our results show that dissipative Floquet dynamics does not have static counterparts even under high-frequency drive in contrast to closed systems. As discussed in the introduction, this can break some properties ensured by Liouvillianity, such as the existence of NESS and the validity of the trajectory method (see Sec. S3 in the Supplemental Material \cite{Sfm} in which we discuss the existence of nonequilibrium steady states and the validity of the trajectory method under Liouvillianity breaking). One of the most promising future directions is Floquet engineering, which is a way to engineer preferable steady states or dynamics of static systems by  FM effective Hamiltonians or Lindbladians \cite{Bukov2015,Oka2019}. Liouvillianity breaking implies that we can engineer steady states or dynamics which are not reproducible in static Markovian systems. In particular, the dynamics essentially different from that of static Markovian systems is considered to be a sign of emergent non-Markovianity in the stroboscopic dynamics, as Ref. \cite{Schnell2020} numerically observed finite memory time in noninteracting Floquet Markovian systems. Thus, it should be interesting to seek for what kind of non-Markovian dynamics appears or how much memory time emerges in interacting Floquet Markovian systems under Liouvillianity breaking with our results.



\textit{Acknowledgment.}---K. M. thanks Y. Michishita for fruitful discussions. This work is supported by JSPS KAKENHI
(Grants No. JP18H01140, JP19H01838, and No. JP20J12930). K. M. is supported by WISE Program, MEXT, and a Research Fellowship for Young Scientists from JSPS. K. T. thanks JSPS for support from Overseas Research Fellowship.


\providecommand{\noopsort}[1]{}\providecommand{\singleletter}[1]{#1}%
%

\clearpage
\renewcommand{\thesection}{S\arabic{section}}
\renewcommand{\theequation}{S\arabic{equation}}
\setcounter{equation}{0}
\renewcommand{\thefigure}{S\arabic{figure}}
\setcounter{figure}{0}

\onecolumngrid
\begin{center}
 {\large 
 {\bfseries Supplemental Materials for \\ ``Liouvillianity breaking in Floquet-Lindblad interacting systems \\ under high-frequency drive'' }}
\end{center}

\begin{center}
Kaoru Mizuta,$\,^{1, \ast}$ Kazuaki Takasan,$\,^{2}$ and Norio Kawakami$\,^1$
\end{center}

\begin{center}
{\small $\,^1$\textit{Department of Physics, Kyoto University, Kyoto 606-8502, Japan} \\
$\,^2$\textit{Department of Physics, University of California, Berkeley, California 94720, USA} \\
(Dated: \today)
}
\end{center}

\vspace{-20pt}


\section{Floquet-Magnus expansions and their properties}\label{S_FM}

\subsection{Form of the $n$-th order Floquet-Magnus effective Lindbladian}
In this section, we describe each order term of the Floquet-Magnus (FM) effective Lindbladian. The FM expansion is a perturbative expansion of the effective Lindbladian $\mathcal{L}_\text{eff}$ [Eq. (2) in the main text] in terms of $||\mathcal{L}(t)||_{\text{op}}/\omega$. Then, each order term $\mathcal{L}_f^{(n)}$ is given by
\begin{eqnarray}
\mathcal{L}_f^{(n)} &=& \sum_\sigma (-1)^{n-\tilde{\theta}(\sigma)} \frac{\tilde{\theta}(\sigma)!(n-\tilde{\theta}(\sigma))!}{n!(n+1)^2 iT}
\int_0^T dt_{n+1} \hdots \int_0^{t_2} dt_1 [\mathcal{L}(t_{\sigma(n+1)}), [\mathcal{L}(t_{\sigma(n)}), \hdots ,[\mathcal{L}(t_{\sigma(2)}),\mathcal{L}(t_{\sigma(1)})]\hdots]], \\
\tilde{\theta}(\sigma) &\equiv& \sum_{m=1}^n \theta(\sigma(m+1)-\sigma(m)), \quad \text{$\theta(x)$: a step function,}
\end{eqnarray}
where $\sigma$ represents the permutation of $\{ 1,2, \hdots, n+1 \}$ \cite{Kuwahara2016s}. We note that each $i$-th order term is composed of $i$-tuple multi-commutators of the Liouvillian $\mathcal{L}(t)$ at different time. The $n$-th order FM effective Lindbladian $\mathcal{L}_f^n$ is defined by the summation up to the $n$-th order, $\mathcal{L}_f^{n} = \sum_{i=0}^n \mathcal{L}_f^{(i)}$. 

In the main text, we consider binary drives described by
\begin{equation}
\mathcal{L}(t)  = \begin{cases}
\mathcal{L}_1 & (0\leq t < \tau) \\
\mathcal{L}_2 & ( \tau \leq t < 2\tau = T),
\end{cases}
\end{equation}
and then the Floquet operator is given by $\mathcal{U}_\text{eff} = \exp (\mathcal{L}_\text{eff} \cdot 2\tau) = \exp (\mathcal{L}_2 \tau) \exp(\mathcal{L}_1 \tau)$. The Baker-Campbell-Hausdorff formula enables the direct calculation of each order term $\mathcal{L}_f^{(i)}$, which results in
\begin{equation}\label{BCH}
\mathcal{L}_f^{(0)} = \frac{1}{2} (\mathcal{L}_1 + \mathcal{L}_2), \quad \mathcal{L}_f^{(1)} = \frac{\tau}{4} [\mathcal{L}_2,\mathcal{L}_1], \quad
\mathcal{L}_f^{(2)} = \frac{\tau^2}{24} [ \mathcal{L}_2 - \mathcal{L}_1, [\mathcal{L}_2, \mathcal{L}_1]], \quad \mathcal{L}_f^{(3)} = \frac{\tau^3}{48} [\mathcal{L}_1,[\mathcal{L}_2,[\mathcal{L}_1,\mathcal{L}_2]]], \quad \hdots
\end{equation}

\subsection{Condition for Liouvillianity}
Ref. \cite{Haddadfarshi2015s} clarified a way to judge the Liouvillianity of the FM effective Lindbladian. Here, we reformulate this using the Frobenius basis $\{ F_j \}$, while the basis for $d$-dimensional square matrices $\mathbb{M}_d$ is not specified in Ref. \cite{Haddadfarshi2015s}. We note that the traceless-property of the Frobenius basis enables us to easily extract an effective Hamiltonian and an effective dissipation from the FM expansions, and to evaluate their Liouvillianity and upper bound, as discussed later. First, we derive the form of the FM expansions, which is the same as Eq. (5) in the main text.
\begin{theorem}\label{Thm_FMeffectiveDoubled}
In the doubled Hilbert space representation, the $n$-th order FM effective Lindbladian $\mathcal{L}_f^n$ is always written in the following form using the Frobenius basis $\{ F_j \}$:
\begin{equation}\label{S_FMeffectiveDoubled}
\mathcal{L}_f^{n} = -i (H^n \otimes I - I \otimes (H^n)^\mathrm{T})
+ \sum_{j,k=1}^{d^2-1} a_{jk}^{n} \left[ F_j \otimes F_k^\ast - \frac{1}{2} ( F_k^\dagger F_j \otimes I - I \otimes F_j^\mathrm{T} F_k^\ast ) \right],
\end{equation}
where  the matrices $H^n \in \mathbb{M}_d$ and $[a_{jk}^{(n)}] \in \mathbb{M}_{d^2-1}$ are hermitian.
\end{theorem}

\textit{Proof.}---The Lindbladian $\mathcal{L}(t)$ [Eq. (1) in the main text] satisfies the following conditions at each time $t$:
\begin{equation}
\mathrm{Tr} (\mathcal{L}(t)[\rho]) = 0, \quad 
(\mathcal{L}(t)[\rho])^\dagger = \mathcal{L}(t)[\rho^\dagger], \quad \,^\forall \rho.
\end{equation}
The first condition represents that the time evolution operator $\mathcal{U}(t)$ is trace-preserving, and the second one represents that $\mathcal{L}(t)$ is hermiticity-preserving. Then, the sum, difference, and product of $\mathcal{L}(t)$ satisfy the same properties. For example,
\begin{equation}
(\mathcal{L}(t_1) \mathcal{L}(t_2) [\rho] )^\dagger = \mathcal{L}(t_1) [ (\mathcal{L}(t_2) [\rho])^\dagger] = \mathcal{L}(t_1) \mathcal{L}(t_2) [\rho^\dagger]
\end{equation}
is obtained. Since $\mathcal{L}_f^n$ is composed of the summation of commutators of $\mathcal{L}(t)$, it possesses the same properties:
\begin{eqnarray}
\mathrm{Tr} (\mathcal{L}_f^{n}[\rho]) &=& 0, \quad \,^\forall \rho, \label{TracePreserve} \\
(\mathcal{L}_f^{n}[\rho])^\dagger &=& \mathcal{L}_f^{n}[\rho^\dagger], \quad \,^\forall \rho. \label{HermiticityPreserve}
\end{eqnarray}

Generic linear operators satisfying Eqs. (\ref{TracePreserve}) and (\ref{HermiticityPreserve}) are written in the form of Eq. (\ref{S_FMeffectiveDoubled}). Although its proof is almost parallel to the one in Ref. \cite{Szczygielski2019s}, we prove this in detail for this paper to be self-contained. The structure theorem \cite{Pillis1967s,Hill1973s} says that a linear operator satisfying Eq (\ref{HermiticityPreserve}) is written in the form of
\begin{equation}
\mathcal{L}_f^{n}[\rho] = \sum_i x_i X_i \rho X_i^\dagger, \quad x_i \in \mathbb{R}, \, X_i \in \mathbb{M}_d.
\end{equation}
We express $X_i \in \mathbb{M}_d$ as $X_i = \sum_{j=1}^{d^2} t_{ij}^{n} F_j$, then we arrive at
\begin{equation}\label{HermiticityA_jk}
\mathcal{L}_f^{n}[\rho] = \sum_{j,k=1}^{d^2} a_{jk}^{n} F_j \rho F_k^\dagger, \quad
a_{jk}^{n} = \sum_i x_i t_{ij}^{n} (t_{ik}^{n})^\ast = (a_{kj}^{n})^\ast.
\end{equation}
The latter equation represents the hermiticity of the matrix $[a_{jk}^{n}]_{j,k=1}^{d^2-1}$. Using the fact $F_{d^2}=I_d/\sqrt{d}$ and defining $M^{n} \equiv (a_{d^2d^2}^{n}/2d)I_d+\sum_{j=1}^{d^2-1} a_{jd^2}^{n} F_j$ result in
\begin{equation}
\mathcal{L}_f^{n}[\rho] = M^{n} \rho + \rho (M^{n})^\dagger + \sum_{j,k=1}^{d^2-1} a_{jk}^{n} F_j \rho F_k^\dagger 
= i [\mathrm{Im} (M^{n}), \rho] +\{ \mathrm{Re} (M^{n}), \rho \}  + \sum_{j,k=1}^{d^2-1} a_{jk}^{n} F_j \rho F_k^\dagger.
\end{equation}
In the last equality, we define two hermitian matrices $\mathrm{Re}(M)=(M+M^\dagger)/2$ and $\mathrm{Im}(M)=(M-M^\dagger)/2i$. Then,
\begin{equation}
\mathrm{Tr} (\mathcal{L}_f^{n}[\rho]) = \mathrm{Tr} [\{ \mathrm{Re} (M^{n}), \rho \}]
+ \sum_{j,k=1}^{d^2-1} a_{jk}^{n} \mathrm{Tr} [F_j \rho F_k^\dagger] \nonumber \\
= \mathrm{Tr} \left[ \left( 2 \mathrm{Re} (M^{n}) + \sum_{j,k=1}^{d^2-1} a_{jk}^{n} F_k^\dagger F_j \right) \rho \right] 
\end{equation}
should be zero regardless of $\rho$ from Eq. (\ref{TracePreserve}). Therefore, $\mathrm{Re} (M^{n})$ is given by
\begin{equation}
\mathrm{Re} (M^{n}) = - \frac{1}{2}  \sum_{j,k=1}^{d^2-1} a_{jk}^{n} F_k^\dagger F_j,
\end{equation}
where the hermiticity of $\mathrm{Re} (M^{n})$ is ensured by Eq. (\ref{HermiticityA_jk}). Finally, by defining $H^n = - \mathrm{Im}(M^n)$ and using the doubled Hilbert space representation, we obtain Eq. (\ref{S_FMeffectiveDoubled}) $\qquad \qquad \square$.

The Liouvillianity of $\mathcal{L}_f^n$ is determined only by $[a_{jk}^n]$. Using the spectral decomposition of the hermitian matrix, $a_{jk}^n = \sum_{i=1}^{d^2-1} \tilde{x}_i \tilde{t}_{ij} (\tilde{t}_{ik})^\ast$ ($\tilde{x}_i \in \mathbb{R}$ and $j, k = 1,2, \hdots, d^2-1$), and defining $\tilde{L}_i = \sum_{j=1}^{d^2-1} \sqrt{|\tilde{x}_i|} \tilde{t}_{ij} F_j$, we can rewrite Eq. (\ref{S_FMeffectiveDoubled}) as follows:
\begin{equation}\label{L_FM_Lindblad_form}
\mathcal{L}_f^{n} = -i (H^n \otimes I - I \otimes (H^n)^\mathrm{T})
+ \sum_{i=1}^{d^2-1} \mathrm{sgn}(\tilde{x}_i) \left[ \tilde{L}_i \otimes \tilde{L}_i^\ast - \frac{1}{2} ( \tilde{L}_i^\dagger \tilde{L}_i \otimes I + I \otimes \tilde{L}_i^\mathrm{T} \tilde{L}_i^\ast ) \right].
\end{equation}
Thus, $\mathcal{L}_f^n$ is a Liouvillian if and only if $[a_{jk}^n]$ is positive-semidefinite ($\tilde{x}_i \geq 0$ for all $i$). Conversely, by expanding $H^n$ by the Frobenius basis as $H^n = \sum_{j=1}^{d^2-1} h_j^n F_j$ (The $j=d^2$ component is irrelevant in the commutator), we obtain
\begin{equation}
\mathcal{L}_f^{n} = -i \sum_{j=1}^{d^2-1} h_j^n (F_j \otimes I - I \otimes F_j^\mathrm{T})
+ \sum_{j,k=1}^{d^2-1} a_{jk}^{n} \left[ F_j \otimes F_k^\ast - \frac{1}{2} ( F_k^\dagger F_j \otimes I + I \otimes F_j^\mathrm{T} F_k^\ast ) \right].
\end{equation}
When we assume that $F_j$ is hermitian (for example, this is satisfied in spin systems in the main text), and then $h_j^n$ is real due to the hermiticity of $H^n$. Multiplying $F_j^\dagger \otimes F_k^T$ ($j,k \neq d^2$) to Eq. (\ref{S_FMeffectiveDoubled}) and taking its trace, we can extract the dissipative components,
\begin{equation}\label{S_a_jk}
a_{jk}^n = \mathrm{Tr} [(F_j^\dagger \otimes F_k^T) \, \mathcal{L}_f^n] = \left< (F_j \otimes F_k^\ast), \mathcal{L}_f^n \right>_\mathrm{F},
\end{equation}
where we have used the traceless-property of the Frobenius basis. In a similar way, under the hermiticity of $F_j$, we obtain
\begin{equation}
\mathrm{Tr} [(F_j \otimes I) \, \mathcal{L}_f^n] = -i h_j^n \mathrm{Tr} I - \frac{1}{2} \sum_{j^\prime k^\prime} a_{j^\prime k^\prime}^n \mathrm{Tr} [F_j F_{k^\prime} F_{j^\prime}] \cdot \mathrm{Tr} I,
\end{equation}
and we can extract the effective Hamiltonian terms $h_j^n$ by
\begin{equation}
h_j^n = \frac{i}{d} \mathrm{Tr} [(F_j \otimes I) \, \mathcal{L}_f^n] + \frac{i}{2} \sum_{j^\prime k^\prime} \mathrm{Tr} [(F_{j^\prime}^\dagger \otimes F_{k^\prime}^T) \, \mathcal{L}_f^n] \cdot \mathrm{Tr} [F_j F_{k^\prime} F_{j^\prime}] .
\end{equation}
We emphasize that Theorem \ref{Thm_FMeffectiveDoubled} is proven only using the fact that $\mathcal{L}_f^{n}$ is given by the integrals of polynomial functions of the Liouvillian $\mathcal{L}(t)$.  Therefore, each $i$-th order term in the FM effective Lindbladian, $\mathcal{L}_f^{(i)}$, is also written in the same form as Eq. (\ref{S_FMeffectiveDoubled}). Other types of high-frequency expansions such as the van Vleck expansion and the Schrieffer-Wolff expansion \cite{Bukov2015s} also satisfy this theorem, and we can check their Liouvillianity in the same way.

\subsection{Liouvillianity of each $i$-th order term $\mathcal{L}_f^{(i)}$}
Here, we derive the propositions  (a) and (b) in the main text, which dictate that a higher order term in the FM effective Lindbladian is not a Liouvillian in general. 
\begin{theorem}\label{Thm_EachOrder}
The zeroth order term of the FM effective Lindbladian, $\mathcal{L}_f^{(0)}$, is always a Liouvillian. On the other hand, for $i \geq 1$, the $i$-th order term $\mathcal{L}_f^{(i)}$ is a Liouvillian if and only if $[a_{jk}^{(i)}] = O$, where the matrix $[a_{jk}^{(i)}]$ is defined by $a_{jk}^{(i)}=\mathrm{Tr} [(F_j^\dagger \otimes F_k^\ast) \, \mathcal{L}_f^{(i)}]$.
\end{theorem}

\textit{Proof.}---As discussed in the last subsection, each $i$-th order term $\mathcal{L}_f^{(i)}$ is always written in the same form as Eq. (\ref{S_FMeffectiveDoubled}), and $\mathcal{L}_f^{(i)}$ is a Liouvillian if and only if $[a_{jk}^{(i)}]=[\mathrm{Tr} [(F_j^\dagger \otimes F_k^\ast) \, \mathcal{L}_f^{(i)}]]$ is positive-semidefinite. Since the zeroth order term is given by the time-average of $\mathcal{L}(t)$,
we obtain
\begin{equation}
a_{jk}^{(0)} = \frac{1}{T} \int_0^T a_{jk}(t) dt.
\end{equation}
The hermitian matrix $[a_{jk}^{(0)}]$ becomes positive-semidefinite since $[a_{jk}(t)]$ is positive-semidefinite, and hence $\mathcal{L}_f^{(0)}$ is always a Liouvillian.
On the other hand, using the fact that each order term $\mathcal{L}_f^{(i)}$ is composed of $i$-tuple comutators, each $i$-th order term is traceless, $\mathrm{Tr} \left[ \mathcal{L}_f^{(i)} \right]= 0$, for $i \geq 1$. We can calculate the trace in another way using Eq. (\ref{S_FMeffectiveDoubled}), and this results in
\begin{equation}
\mathrm{Tr} \left[ \mathcal{L}_f^{(i)} \right] = - \frac{1}{2} \sum_{j,k=1}^{d^2-1} a_{jk}^{(i)} (\mathrm{Tr} [F_j^\dagger F_k \otimes I ] + \mathrm{Tr} [I \otimes F_j^\mathrm{T} F_k^\ast])
= - d \cdot \sum_{j=1}^{d^2-1} a_{jj}^{(n)} = -d \cdot \mathrm{Tr} \left( [a_{jk}^{(i)}] \right).
\end{equation}
Therefore, $[a_{jk}^{(i)}]$ is also traceless, and hence the summation of the eigenvalues of $[a_{jk}^{(i)}]$ is zero. Since all of the eigenvalues of hermitian positive-semidefinite matrices cannot be negative, $[a_{jk}^{(i)}]$ is positive-semidefinite if and only if $[a_{jk}^{(i)}]=O$. Using the condition for Liouvillianity, we complete the proof of the theorem. $\qquad \square$

Importantly, this theorem is derived from that the zeroth order $\mathcal{L}_f^{(0)}$ is the time-average of $\mathcal{L}(t)$, and that the higher order terms $\mathcal{L}_f^{(i)}$ are composed of commutators. Thus, this theorem also holds for other types of high-frequency expansion with the same properties, such as the van Vleck expansion and the Schrieffer-Wolff expansion.

\subsection{Upper bound of dissipative terms in the FM effective Lindbladian}
We derive the upper bound of the matrix elements $a_{\vec{j}\vec{k}}^{(i)}$ in few-body or many-body systems with local interactions. Before discussing the result, we rigorously define the locality and the extensiveness of dissipative systems dominated by a Liouvillian $\mathcal{L}$, the time-independent version of Eq. (1) in the main text.  We call a Liouvillian $\mathcal{L}$ being $k$-local when its Hamiltonian $H$ and Lindblad operator $L_i$ include at-most $k$-body and $(k/2)$-body interactions respectively. Let us define $\mathcal{L}|_X$ by the terms in $\mathcal{L}$, which nontrivially act on the domain $X$ in the doubled Hilbert space representation. Note that the number of the sites becomes doubled in the doubled Hilbert space representation, and that a domain $X$ is a subset of $\{ 1,2, \hdots, 2L \}$. Then, a Liouvillian $\mathcal{L} = \sum_X \mathcal{L}|_X$ is called $J$-extensive when
\begin{equation}
\sum_{X: X \owns i} || \mathcal{L}|_X ||_\text{op} \leq J, \quad \,^\forall i \in \{1,2,\hdots,2L\}
\end{equation}
is satisfied. The left hand side means the maximal complex energy at each site $i$, and the extensiveness represents the complex energy per site $\sim || \mathcal{L}||_\text{op} / 2L$ is bounded by $J$. We note that these definitions are the extended versions of those in Refs. \cite{Kuwahara2016s,Mori2016s} generalized to dissipative cases. With their rigorous definitions, we obtain the following result on the upper bound of $a_{\vec{j}\vec{k}}^{(i)}$.

\begin{theorem}
We consider an $L$-site system where each site has $f$-degrees of freedom, and suppose that its Liouvillian $\mathcal{L}(t)$ is $k$-local and $J$-extensive at every time $t$. Then, the dissipative terms of each $i$-th order term $\mathcal{L}_f^{(i)}$, represented by $[a_{\vec{j}\vec{k}}^{(i)}]$, has the following upper bound:
\begin{equation}
\left|  a_{\vec{j}\vec{k}}^{(i)}  \right| \leq \frac{(2kJT)^i}{i+1} J \cdot i! \cdot f^L.
\end{equation}
\end{theorem}

\textit{Proof.}---We consider $a_{\vec{j}\vec{k}}^{(i)}$ for some fixed $\vec{j}, \vec{k} \neq \vec{0}$, and let $X$ be a domain where $F_{\vec{j}} \otimes F_{\vec{k}}^\ast$ nontrivially acts in the $2L$-site system. Let us define $A_{\vec{j}\vec{k}}^{(i)}$ by
\begin{equation}
A_{\vec{j}\vec{k}}^{(i)} = \sum_{\vec{j}^\prime, \vec{k}^\prime} a_{\vec{j}^\prime  \vec{k}^\prime}^{(i)} \left( F_{\vec{j}^\prime} \otimes F_{\vec{k}^\prime}^\ast \right),
\end{equation}
where $\sum_{\vec{j}^\prime, \vec{k}^\prime}$ represents the summation over $\vec{j}^\prime, \vec{k}^\prime$ such that $F_{\vec{j}^\prime} \otimes F_{\vec{k}^\prime}^\ast$ nontrivially acts only on the domain $X$ (there exist at-most $(f^2-1)^{|X|}$ terms). Then, $A_{\vec{j}\vec{k}}^{(i)}$ is the unique term nontrivially acting just on $X$ in $\mathcal{L}_f^{(i)}$, and hence we obtain
\begin{equation}\label{S_A_jk_bound}
||A_{\vec{j}\vec{k}}^{(i)}||_\text{op} \leq J^{(i)},
\end{equation}
where  $J^{(i)}$ is the extensiveness of the $i$-th order term $\mathcal{L}_f^{(i)}$ from the definition of the extensiveness. When we assume the $k$-locality and the $J$-extensivesness of the Lindbladian $\mathcal{L}(t)$, it is known that $J^{(i)}$ is bounded as follows (See Lemma 5 in Ref. \cite{Kuwahara2016s}):
\begin{equation}\label{S_J_bound}
J^{(i)} \leq \frac{(2kJT)^i}{i+1} J \cdot i!.
\end{equation}
Let us define the Frobenius norm $|| \quad ||_\text{F}$  for square matrices by $|| A ||_\text{F} = \sqrt{\left< A, A \right>_\text{F}}$, and then, using the Schwartz inequality, we arrive at
\begin{equation}
\left| a_{\vec{j}\vec{k}}^{(i)} \right| = \left| \left< \left(F_{\vec{j}}
 \otimes F_{\vec{k}} \right), A_{\vec{j}\vec{k}}^{(i)} \right>_\text{F} \right| 
\leq \left|\left| \left(F_{\vec{j}}^\dagger \otimes F_{\vec{k}}^T \right) \right|\right|_\text{F} \cdot
\left|\left| A_{\vec{j}\vec{k}}^{(i)}  \right|\right|_\text{F}.
\end{equation}
Using the relation $||A||_\text{F} \leq \sqrt{\text{rank}A} \cdot ||A||_\text{op}$ and Eqs. (\ref{S_A_jk_bound}) and (\ref{S_J_bound}), we obtain
\begin{equation}
\left| a_{\vec{j}\vec{k}}^{(i)} \right|  \leq \sqrt{\text{rank} \left( A_{\vec{j}\vec{k}}^{(i)} \right)} \cdot \left|\left| A_{\vec{j}\vec{k}}^{(i)}  \right|\right|_\text{op} 
\leq \frac{(2kJT)^i}{i+1} J \cdot i! \cdot f^L,
\end{equation}
in which $f=2$ reproduces the result Eq. (17) in the main text. $\qquad \qquad \square$

\section{Liouvillianity breaking in interacting systems: another model}\label{S_Models}
In this section, we provide another example for interacting cases. As we will see later, this model shares some properties concerning Liouvillianity breaking with the interacting model $\mathcal{L}_C(t)$ in the main text. Discussion in this section will help us understand that Liouvillianity breaking in interacting systems is not limited to some specific models.

We consider an $L$-site Ising spin chain driven by the following time-periodic Liouvillian:
\begin{equation}
\mathcal{L}_D(t) \rho = \begin{cases}
-i J_x \sum_l [\sigma_l^1 \sigma_{l+1}^1, \rho] \equiv \mathcal{L}_{D1} \rho & ( 0 \leq t < \tau )\\
\gamma \sum_l (\sigma_l^{-} \rho \sigma_l^{+} - \frac{1}{2} \{ \sigma_l^{+} \sigma_l^{-}, \rho \} ) \equiv \mathcal{L}_{D2} \rho & ( \tau \leq t < 2\tau=T ),
\end{cases}
\end{equation}
with $\sigma_l^{\pm} = (\sigma_l^1 \pm i \sigma_l^2)/2$.
The first step represents the nearest neighbor Ising interactions in $x$-direction, and the second one represents a noise which makes the down spin state preferable. The calculation up to the first order results in
\begin{eqnarray}
\mathcal{L}_f^1 &=& \frac{\mathcal{L}_{D1}+\mathcal{L}_{D2}}{2} - \frac{\gamma J_x \tau}{8} \sum_l \left\{ i \sigma_l^1 \otimes \sigma_l^3 (\sigma_{l-1}^1+\sigma_{l+1}^1) -i \sigma_l^3 (\sigma_{l-1}^1+\sigma_{l+1}^1) \otimes \sigma_l^1 \right\} \nonumber \\
&\qquad& - \frac{\gamma J_x \tau}{8} \sum_l \left\{ \sigma_l^2 \otimes \sigma_l^3 (\sigma_{l-1}^1 + \sigma_{l+1}^1) - \sigma_l^3 (\sigma_{l-1}^1 - \sigma_{l+1}^1) \otimes \sigma_l^2  \right\} \nonumber \\
&\qquad& - \frac{\gamma J_x \tau}{8} \sum_l \left\{ (\sigma_l^1 \sigma_{l+1}^2 + \sigma_l^2 \sigma_{l+1}^1) \otimes I + I \otimes (\sigma_l^1 \sigma_{l+1}^2 - \sigma_l^2 \sigma_{l+1}^1 )\right\},
\end{eqnarray}
\begin{equation}
[a_{\vec{j}\vec{k}}^{1}] =  \left[ \bigoplus_{l=1}^L  \gamma \cdot 2^{L-3} \left( \begin{array}{cccc}
2 & 2 i & -i (J_x \tau) & -i (J_x \tau)  \\
-2 i & 2 & - (J_x \tau) & - (J_x \tau) \\
i (J_x \tau) &  - (J_x \tau) & 0 & 0 \\
i (J_x \tau)  & - (J_x \tau) & 0 & 0
\end{array} \right)_l
\right] \oplus O_{4^L-4L-1}.
\end{equation}
The basis for the nontrivial $4 \times 4$ matrices in this equation is composed of $\vec{j}=(\hdots,0, j_l=1,0,\hdots),(\hdots,0,j_l=2,0,\hdots),(\hdots,0,1,j_l=3,0,\hdots),(\hdots,0,j_l=3,1,0,\hdots)$. The first order terms proportional to $(J_x \tau)$ come from three-body terms of $\mathcal{L}_f^{1}$ in the doubled Hilbert space representation, which are just manifestation of the propagating local interactions via the FM expansions. This spreading structure of the matrix $[a_{\vec{j}\vec{k}}^{n}]$ is common in generic interacting systems as discussed in the main text, leading to the results that $[a_{\vec{j}\vec{k}}^1]$ always has a negative eigenvalue $\gamma \cdot 2^{L-2} (1 - \sqrt{1+(J_x \tau)^2})$. Therefore, this interacting model also hosts Liouvillianity breaking of the FM effective Lindbladians under any finite drive, as well as the model in the main text. We note that, since the negative eigenvalue of $[a_{\vec{j}\vec{k}}^1]$ is small compared to the truncation order, the Liouvillianity breaking up to the first order can be removed in these models by using the same way for the noninteracting models in the previous section. However, when we consider higher order FM effective Lindbladians, Liouvillianity breaking can be no longer removed in general as well.

\section{Existence of NESS and Breakdown of trajectory method}
Liouvillianity breaking of the FM effective Lindbladian implies that we cannot use conventional theories for static Markovian systems brought by Liouvillianity. We do not know whether individual theories constructed in static systems so far are valid even in the absence of Liouvillianity, but we show that it can possibly break the two important generic notions, the existence of NESS and the validity of the trajectory method.

Let us consider an effectively static system driven by the FM effective Lindbladian $\mathcal{L}_f^n$. Nonequilibrium steady states (NESS) exist if and only if $\mathcal{L}_f^n$ has at least one zero-eigenvalue and all the eigenvalues of  $\mathcal{L}_f^n$ have nonpositive real parts. Then, the right eigenstates with zero eigenvalues are called NESS. The existence of NESS is ensured under Liouvillianity. On the other hand, $\mathcal{L}_f^n$ is written as
\begin{equation}
\mathcal{L}_f^n \rho = -i [H^n, \rho] + \sum_i s_i \left( L_i^n \rho L_i^{n\dagger} - \{ L_i^{n\dagger} L_i^n, \rho \} \right) , \quad s_i = \pm 1,
\end{equation}
and some of $\{ s_i \}$ become negative if Liouvillianity is broken [See Eq. (\ref{L_FM_Lindblad_form})]. From this representation, $(\mathcal{L}_f^n)^\dagger I_d = O$ ($I_d$: the $d$-dimensional identity matrix) is satisfied, indicating that $I_d$ is the left eigenstate of $\mathcal{L}_f^n$. Thus, $\mathcal{L}_f^n$ always has at least one right eigenstate $\rho_0$ with zero eigenvalue. However, all the eigenvalues of $\mathcal{L}_f^n$ do not necessarily have nonpositive real parts, and hence the state $\rho_0$ does not always represent NESS. Therefore, $\mathcal{L}_f^n$ which breaks Liouvillianity does not ensure the existence of NESS in general.

In our models $\mathcal{L}_\alpha (t)$ ($\alpha=A,B,C,D$), NESS becomes ill-defined under  $\mathcal{L}_f^n$ ($n \geq 1$) when the frequency is comparable to the energy scale, although such an anomalous effect is not physically accessible. However, we expect that, when the NESS of the Liouvillian $\mathcal{L}_f^0$ is degenerated or the Liouvillian gap $\Delta=\min \{ \mathrm{Re} \lambda \neq 0 \, | \text{ $\lambda$ : eigenvalue of $\mathcal{L}_f^0$} \}$ is small enough, some anomalous behaviors can be observed in physically relevant regimes, which will be left for future work. 


Next, we discuss the validity of trajectory method, with which we can efficiently calculate the nonequilibrium dynamics \cite{Daley2014s}.  In static Liouvillian systems, the Lindblad equation is rewritten as
\begin{equation}\label{LindbladStatic}
\partial_t \rho = -i (H_\text{eff} \rho - \rho H_\text{eff}^\dagger) + \sum_i L_i \rho L_i^\dagger
\end{equation}
with a non-hermitian Hamiltonian $H_\text{eff}=H - (i/2) \sum_i L_i^\dagger L_i$. A single trajectory dynamics is a stochastic dynamics composed of non-unitary time evolution under $H_\text{eff}$ and quantum jumps by $L_i$. Let us assume the initial state $\rho_0 = \ket{\psi_0}\bra{\psi_0}$ and consider the dynamics of $\ket{\psi_0}$ for infinitesimal duration $\delta t$. Up to the first order of $\delta t$, the state is stochastically updated by $\exp (-i H_\mathrm{eff} \delta t) \ket{\psi_0}/\sqrt{1-p}$ with the probability $1-p$ (non-hermitian dynamics) or by $L_i \ket{\psi_0} / \sqrt{p_i}$ with the probability $p_i$ (quantum jumps). Here, the propabilities $p_i$ and $p$ are given by $p_i = \braket{\psi_0 | L_i^\dagger L_i | \psi_0}$ and $p=1-\sum_i p_i$ respectively. A series of states $\ket{\psi(t)}$ obtained by repeating this procedure $m$ times up to $t=m \delta t$ is called a trajectory. By taking the statistical ensemble of $\ket{\psi(t)}\bra{\psi(t)}$ over many trajectories with small $\delta t$, we can reproduce the density operator $\rho(t)$ obeying the Lindblad equation Eq. (\ref{LindbladStatic}). We note that all the eigenvalues of $H_\text{eff}$ have nonpositive imaginary parts, indicating that the non-unitary time evolution by $H_\text{eff}$ is always lossy. The lost probability due to this non-unitary dynamics corresponds to the probabilities of quantum jumps by $L_i$.

On the other hand, if $\mathcal{L}_f^n$ breaks Liouvillianity, the corresponding non-hermitian Hamiltonian becomes $H_\mathrm{eff} = H -(i/2) \sum s_i L_i^{n\dagger} L_i^n$ ($s_i= \pm 1$) where some of $\{ s_i \}$ are $-1$. Thus, $H_\mathrm{eff}$ can have eigenvalues with positive imaginary parts, and then stochastic dynamics composed of the non-hermitian Hamiltonian time evolution and quantum jumps becomes ill-defined (some of the probabilities $p_i$ become negative). This represents the breakdown of the trajectory method in the absence of Liouvillianity for $\mathcal{L}_f^n$.

\providecommand{\noopsort}[1]{}\providecommand{\singleletter}[1]{#1}%

\end{document}